\documentclass[aps,prl,twocolumn,showpacs]{revtex4}
\usepackage[T1]{fontenc}
\usepackage[latin1]{inputenc}
\usepackage{graphics}
\usepackage{amsmath}
\usepackage{amssymb}
\usepackage{ulem}

\newcommand{\beq}[1][]{\begin{equation}\label{#1}}
\newcommand{\eeq}{\end{equation}}

\newcommand{\T}[1]{\tilde{#1}}
\newcommand{\eps}{\varepsilon}
\renewcommand{\phi}{\varphi}
\renewcommand{\vec}[1]{\boldsymbol{#1}}



\begin{document}

\title{Scaling properties of $q$-breathers in nonlinear acoustic lattices}



\author{O.~I.~Kanakov$^1$, S. Flach$^2$, M.~V.~Ivanchenko$^1$ and
K.~G.~Mishagin$^1$}

\affiliation{$^1$ Department of Radiophysics, University of Nizhny Novgorod,
Gagarin Avenue, 23, 603950 Nizhny Novgorod, Russia
\\
$^2$ Max-Planck-Institut f\"ur Physik komplexer
Systeme, N\"othnitzer Str. 38, D-01187 Dresden, Germany }

\begin{abstract}
Recently $q$-breathers - time-periodic solutions which localize
in the space of normal modes and maximize the energy density for
some mode vector $q_0$ - were obtained for finite nonlinear
lattices. We scale these solutions together with the size of the
system to arbitrarily large lattices. We generalize previously
obtained analytical estimates of the localization length of
$q$-breathers. The first finding is that the degree of
localization depends only on intensive quantities and is size
independent. Secondly a critical wave vector $k_m$ is
identified, which depends on one effective nonlinearity
parameter. $q$-breathers minimize the localization length at
$k_0=k_m$ and completely delocalize in the limit $k_0
\rightarrow 0$.

\end{abstract}


\maketitle

\section{Introduction}

Spatially extended nonlinear Hamiltonian systems serve as starting
models for the study of excitations in many branches in physics,
e.g. anharmonic vibrations of crystal lattices, mesoscopic and
nanoscopic systems, molecules, but also of electromagnetic,
acoustic and other waves in nonlinear media, to name a few. They
have been studied over many decades in order to understand such
intriguing material properties as heat conductivity, thermal
expansion, turbulence, confinement of light, but also general
mathematical aspects such as thermalization, mode-mode
interactions, etc. While in any realistic situation damping and
energy input have to be considered as well, these dissipative
effects are often weak enough to allow the observation of the
underlying Hamiltonian excitations.

Recently it was shown \cite{QB1}, that a one-dimensional anharmonic atomic
chain allows for exact time-periodic solutions which localize exponentially
in the space of normal modes and have their maximum energy on a mode with
mode number $q_0$.
A $q$-breather, being periodic in time,
can be viewed as one normal mode which is dressed by several other
normal modes in a neighbourhood of $q_0$ and has an infinite lifetime.
The existence and properties of these $q$-breathers allowed
to explain all major ingredients of the famous Fermi-Pasta-Ulam
problem (FPU) \cite{fpu,Ford,chaosfpu} in a clear and constructive way. The FPU problem
concerns the nonequipartition of normal mode energies on time scales which
can be many orders of magnitude larger than the characteristic vibration periods.
The key ingredient for the construction of $q$-breathers is a finite nonlinear
system \cite{QB1}. This is a very general condition and may apply to many other systems
as well. It has been recently successfully tested by considering FPU models
with lattice dimension $d=2,3$ \cite{QB23D} and also discrete nonlinear
Schr\"odinger models (DNLS)
in various lattice dimensions \cite{QB-DNLS}.
Further studies also revealed
the persistence of $q$-breathers in thermal equilibrium \cite{QB1},
which shows their relevance for statistical properties of extended systems.
Previous studies of the FPU problem suggested that the effect of
nonequipartition will disappear for large system sizes \cite{Ford,chaosfpu}.
That seems to imply a disappearance of $q$-breathers in that limit.
Here we show that $q$-breathers persist and have invariant properties
for large system sizes. These results are particularly important because
they apply to macroscopic systems.

Consider a generic model of a $D$-dimensional nonlinear lattice
of size $N_1\times \dots \times N_D$, defined by a Hamiltonian
\beq[H]
H=\sum_{\vec{n}} \left[
\frac{p_{\vec{n}}^2}{2}+U(x_{\vec{n}})+ \sum_{l=1}^D
V(x_{\vec{n}+\vec{e}_{l}}-x_{\vec{n}}) \right]
\eeq
where
$x_{\vec{n}}$, $p_{\vec{n}}$ are canonical variables.
$U(x)$ and $V(x)$ are anharmonic on-site and interaction
potentials, respectively. Their Taylor expansion around $x=0$
starts with quadratic terms. $\vec{n}=\{n_1,\dots,n_D\}$ is a
$D$-dimensional lattice vector with $n_{l}=\overline{1,N_{l}}$.
$\vec{e}_{l}$ denotes a unitary lattice vector along the
dimension $l$.
Note that $x_{\vec{n}}=0$ is
an equilibrium state of the system.
We will consider the case of fixed boundary conditions (BC).
We have also studied free and
periodic BC with similar results.

This Hamiltonian can be expressed in terms of normal modes
$P_{\vec{q}}$, $Q_{\vec{q}}$ of the
linearized problem, obtained by skipping all anharmonic terms in the potentials:
\beq[Hmode]
H=\sum_{\vec{q}}E_{\vec{q}}
    +H^{int}(Q_{\vec{q}})\;,\;
    E_{\vec{q}}=\frac{1}{2}\left( P_{\vec{q}}^2+\omega_{\vec{q}}^2 Q_{\vec{q}}^2 \right)
\eeq
where $E_{\vec{q}}$ is the energy of a given normal mode and
$H^{int}$ is the mode interaction part of the Hamiltonian which appears
due to anharmonicity. The integer components of the mode vector $\vec{q}$ enumerate
the modes.
That is a class of models for which exact $q$-breather
solutions may exist \cite{QB1,QB23D}.
Such solutions are time-periodic and the
normal mode energies are exponentially localized around a mode vector $\vec{q}_0$.

We search for a way of scaling a given solution
$Q_{\vec{q}}(t)$ of a system \eqref{Hmode} of size
$N$ to a
solution $\T Q_{ \T{\vec{q}} }(t)$ of a system of scaled size.
It consists of scaling
the values of mode variables $\sqrt{r}$ times and scaling their indices $r$
times, filling the gaps with zeros:
\beq[scalesol1]
{\Tilde Q}_{\T q_{l}}(t)=
\begin{cases}
\sqrt{r} Q_{q_{l}}(t), & \T q_{l}=r q_{l}
\\
0, & \T q_{l} \neq r q_{l}
\end{cases},
\eeq
where we omitted all mode indices except the $l$-th component.
For fixed BC $q_l=\overline{1,N_l}$, and the scaled system size is assumed
$\T N_l+1=r(N_l+1)$.
The phase space of the scaled system then possesses an invariant subspace (also
coined a bush of modes \cite{bushes}).  This procedure can be applied to
construct a solution to a system of arbitrarily large size, increasing $r$. In
order to identify the  applicability of (\ref{scalesol1}),
we will use a renormalization in
real space which is equivalent to (\ref{scalesol1}).

\begin{figure}
{
\centering
\resizebox*{0.90\columnwidth}{!}{\includegraphics{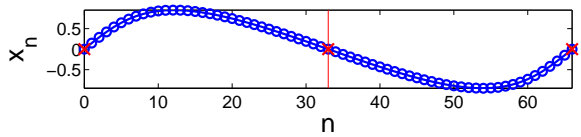}}
}
{\caption{Constructing a solution in a chain of double size in real space. Boundaries
and the additional lattice site in the center are marked with crosses. Initial size
is N=32. The momentary displacements versus lattice site are shown. (color online)}
 \label{fig_scale}}
\end{figure}

Consider $D=1$ in  \eqref{H}.
The equations of motion read
\begin{multline}\label{chain}
\ddot x_n=f(x_n)+\phi(x_{n+1}-x_n)-\phi(x_n-x_{n-1})\;.
\end{multline}
Here $f(x)=-U'(x)$, $\phi(x)=V'(x)$ and $ \quad
n=\overline{1,N}$. For fixed BC $x_0=x_{N+1}=0$. We further
assume an odd restoring force  $f(x)=-f(-x)$. Then (\ref{chain})
is invariant under the following combined parity and sign
reversal symmetry: $x_n \rightarrow -{x}_{N+1-n}$. The
transformation (\ref{scalesol1})  is given by an alternation of
spatial blocks,  obtained from the previous by parity and sign
reversal transform. The blocks are separated by additional
nonexcited lattice sites (see Fig.~\ref{fig_scale}):
\[
\T{x}_n(t)=\{{x}_1(t),\dots,{x}_{N}(t),0,
-{x}_{N}(t),\dots,-{x}_1(t),0,\dots\}.
\]

It is straightforward to observe that if
${x}_{n}(t)$ is a solution to the initial system, then
$\T{x}_{n}(t)$ is a solution to
the scaled-size system.
The scaling rule \eqref{scalesol1} is thus confirmed for 1D chains
\eqref{chain}.

We generalize the above results to higher lattice dimensions.
The transformation to mode variables is a
superposition of 1D transforms.
Then, the transformation \eqref{scalesol1} along a lattice direction
$l$ corresponds to the real-space transform already discussed for the $D=1$ case.
It yields a solution to a scaled-size system if both $f(x)$ and $\phi(x)$ are odd functions.

Given a $q$-breather solution for the original finite system, we can thus
scale the $q$-breather solution to larger system sizes.
Its total energy is scaled in all cases like $\T{E}=r {E}$, which is ensured by the
block structure of the scaling
and the local structure of the coupling in the Hamiltonian
\eqref{H}. The time-dependent mode energies
$E_q$ are transformed as
$\T{E}_{\T q}(t)=r {E}_q(t)$ for $\T q=rq$, and $\T{E}_q=0$ for other $q$.
Introducing the energy density as $\eps=E/(N+1)$  and wave number $k=\pi q/(N+1)$,
it is straightforward to observe that
the scaling procedure leaves the energy density and the wave number (vector)
of a $q$-breather invariant. Together with the rigorous proof of existence of
$q$-breathers for finite systems \cite{QB1} we arrive at a rigorous proof of
existence of these excitations for infinite system sizes, with proper scaling
and under certain restrictions for the potential functions. Since the scaled
solutions are embedded on mode bushes \cite{bushes}, the question arises whether
$q$-breathers with other (or any) values of $k_0$ exist in large systems as well.
The fact that the scaling preserves the localization properties
of the scaled excitations suggests a positive answer. Below we will address this
question. We also note that we needed certain symmetry properties of the potentials
for the scaling to work. It however does not imply that for cases with less symmetries
$q$-breathers do not exist.

The localization properties of $q$-breathers in the normal mode space have been
obtained analytically using asymptotic expansions for FPU-models in various lattice
dimensions. For $f=0$ and
$\phi=x+\beta x^3$, the result from \cite{QB1,QB23D}
is expressed in total energies and mode numbers
$E_{(2n+1)q_0}= \lambda^n E_{q_0}$ with
$\sqrt{\lambda}=3\beta E_{q_0} (N+1)/(8\pi^2 q_0^2)$ for $D=1$ and $q_0 \ll N$.
We conclude that it
must be possible to substitute densities and wave numbers instead,
and obtain an expression which is independent of the actual system size.
Indeed the outcome of our substitution  is written in the following way:
\beq[betloc]
\ln \eps_k = \left(\frac{k}{k_0}-1 \right) \ln \sqrt{\lambda} + \ln \eps_{k_0}, \quad
\sqrt{\lambda}=\frac{3\beta}{2^{2+D}} \frac{\eps_{k_0}}{k_0^2}.
\eeq
Note that these results hold actually for $D=1,2,3$ \cite{QB23D}.
We observe the same scaling properties of (\ref{betloc})
as suggested by the real space renormalization procedure from above.
Equation (\ref{betloc}) implies a smooth dependence of
$\lambda$ on $\eps_{k_0}$ and $k_0$.
We thus test whether the scaling can be observed,
when the size of the system takes
different values, and the new wave number $\tilde{k}_0$ is
chosen to be the nearest one to the original $k_0$ value.
\begin{figure}
{
\centering
\resizebox*{0.90\columnwidth}{!}{\includegraphics{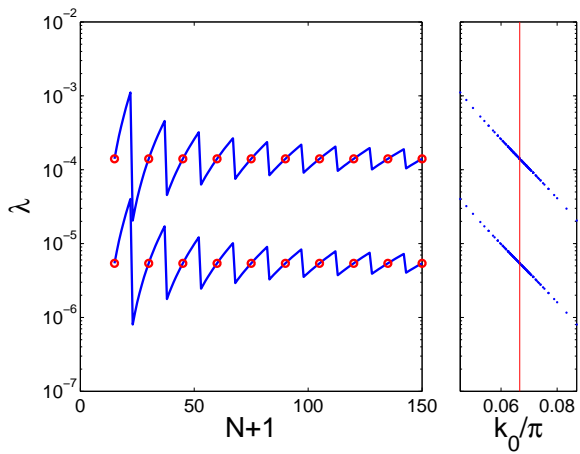}}
}
{\caption{Dependence of the localization parameter $\lambda$
on the system size $N$ of a $q$-breather (left panel) for the
$\beta$-FPU chain at constant
energy density, $\eps=4\cdot 10^{-4}$ and $2\cdot 10^{-3}$ (for lower and upper
curves, respectively) and $\beta=1$. $q_0$ is chosen to be
the nearest integer to $k_0 (N+1)/\pi$, $k_0=\pi/15$.
Open circles mark system sizes $N+1=r(N_0+1)$, $N_0=15$,
for which the prediction is that $\lambda$ should be
invariant on the integer $r$. The right panel shows the
smooth dependence of the measured $\lambda$ values on the actual wave numbers
$k_0$ used in the left panel. (color online)
}
 \label{fig_N}}
\end{figure}
We compute $q$-breathers for that model with $D=1$
for various system sizes starting with $N_0=15$ and plot the numerically
found $\lambda$
as a function of $N$ in the left panel in Fig.\ref{fig_N}.
The mode number $q_0$ is chosen to be
the nearest integer to $k_0 (N+1)/\pi$, $k_0=\pi/15$.
We obtain $\lambda$ from the ratio of the energy densities
$\eps_{5q_0} / \eps_{3q_0}$.
First we observe that $\lambda$ is independent on $N$ when
$N+1=r(N_0+1)$. Secondly we observe fluctuations of $\lambda$ around
a mean value for other values of $N$ due to the fact that
for these system sizes the closest wave number to $k_0$ will nevertheless
be slightly different. These deviations decrease with increase of the
system size, and thus the fluctuation amplitude in Fig.\ref{fig_N}
decreases as well. The piecewise smooth curves show
that $\lambda$ depends smoothly on $k_0$, since by construction
we probe $\lambda$ with wavenumbers slightly varying around $k_0$.
That is nicely confirmed in the right panel in Fig.\ref{fig_N}.

Let us analyze the $k$-dependence of (\ref{betloc}) for $D=1$.
For $q$-breathers at
fixed average energy density $\eps$ it follows within the approximation of exponential
localization that $\eps_{k_0} = (1-\lambda)\eps$.
Together with the definition of $\lambda$ in (\ref{betloc}) we find,
that the inverse localization
length in $k$-space is given by the absolute value of the slope $S$:
\begin{equation}
S= \frac{1}{k_0} \ln \sqrt{\lambda}\;,\;
\sqrt{\lambda} = \frac{\sqrt{1+4\nu^4/k_0^4}-1}{2\nu^2/k_0^2}\;,\;
\nu^2=\frac{3\beta}{8}\eps\;.
\label{slope}
\end{equation}
$\lambda$ depends both on $k_0$ and $\eps_{k_0}$.
$S$
vanishes for $k_0 \rightarrow 0 $ and has its largest absolute value
$max(|S|) \approx 0.7432/\nu$  at
$k_{min}\approx 2.577 \nu$.
For a fixed effective nonlinearity parameter $\nu$
the $q$-breather with $k_0=k_{min}$ shows the strongest localization.
Especially for $k_0 \rightarrow 0$ the $q$-breather delocalizes completely.
With increasing $\nu$,
the localization length of the $q$-breather for $k_0=k_{min}$
increases. For $k_0 \gg \nu$ it follows $S \approx 2/k_0 \ln (\nu/k_0)$
and for $k_0 \ll \nu$ we find $S \approx -k_0/(2\nu^2)$.

\begin{figure}
{
\centering
\resizebox*{0.90\columnwidth}{!}{\includegraphics{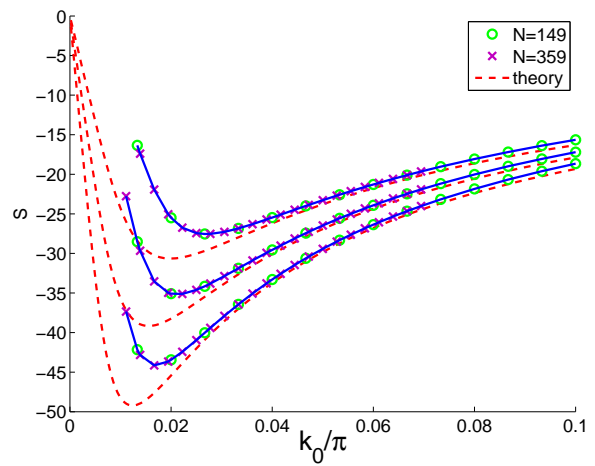}}
}
{\caption{The slope $S$ as a function of $k_0$ for the $\beta$-FPU chain
and three different energy densities
$\eps = 6.08\cdot 10^{-4},9.6\cdot 10^{-4},1.57\cdot 10^{-3}$
and $\beta=1$
(dashed curves, from bottom to top). Symbols and eye-guiding solid lines:
estimate of the slope from numerical computations of $q$-breathers for $N=149$
and $N=359$. (color online)
}
 \label{slopefig}}
\end{figure}

We plot in Fig.\ref{slopefig} the dependence
$S(k_0)$ for the $\beta$-FPU chain at three
different energy densities.
If $\nu$ or $N$ are small enough, then the first nonzero $k_0$ value will appear for
$k_0 > k_{min}$. Increasing $\nu$ or $N$ we shift some allowed low lying
$k$ values to the left of the minimum  $k_0 < k_{min}$.
For very large systems (dense filling of the
$x$-axis in Fig.\ref{slopefig} with allowed $k_0$ values)
we thus expect that
among them there is an optimum wavelength which provides
strongest localization.
\begin{figure}
{ \centering
\resizebox*{0.90\columnwidth}{!}{\includegraphics{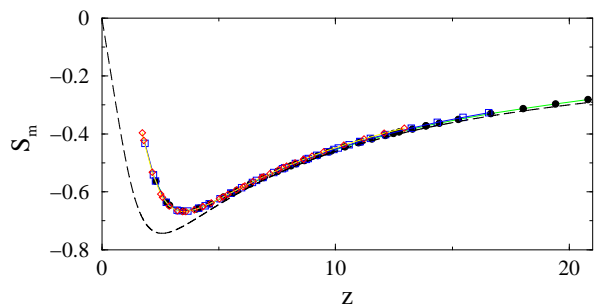}} }
{\caption{ The master slope function $S_m(z)$ (dashed line).
Different symbols correspond to the numerically obtained slopes
from Fig.\ref{slopefig} which are scaled accordingly and fall on a
single curve. (color online) }
 \label{masterslope}}
\end{figure}
We test our prediction by computing the slope for various
$q$-breathers of the $\beta$-FPU chain. The results are shown in Fig.\ref{slopefig} (symbols).
\begin{figure}
{
\centering
\resizebox*{0.90\columnwidth}{!}{\includegraphics{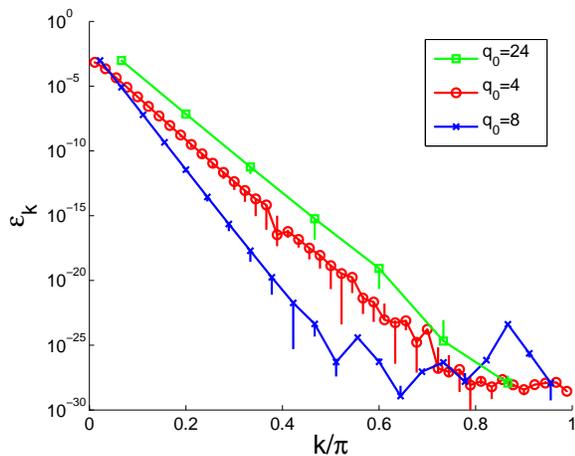}}
}
{\caption{
The profiles of $q$-breather solutions for $N=359$, $\beta=1$,
$\eps=9.6\cdot 10^{-4}$ and three different values of $q_0$
which yield $k_0/\pi=0.011,0.022,0.067$ and correspond to
the most left point on the middle curve in Fig.\ref{slopefig},
its minimum and a point to the right of it.
Symbols denote the mode energies at the moment of vanishing coordinates
$x_l=0$. The vertical bars denote the range of mode energy values
taken during one period of the $q$-breather.  (color online)
}
 \label{qbreatherprofile}}
\end{figure}
We nicely observe an optimum value of $k_0$ for localization.
Increasing the nonlinearity parameter (by either
increasing the nonlinearity strength or the energy density) the critical wavenumber
is shifted further away from the edge of the spectrum and gets shallower, as predicted.
Deviations from the theoretical curves for small $k_0$ are due to
strong nonlinear corrections to our estimates, while deviations at larger
$k_0$ are finite size corrections to the analytical estimates.
Note the smooth dependence of $S$ on $k_0$ which does not depend
on the system size, thus confirming the scaling results from above.
We plot in Fig.\ref{masterslope}  the single master slope function 
$S_m(z)=\nu S$ with $z=k_0/\nu$ and scale
all numerically obtained slopes as well.
The numerical data all condense on one single curve
at small $k_0$.
Thus higher order corrections to the decay law do not alter the
scaling properties of $q$-breathers in the limit of small $k_0$.

The profiles of three $q$-breather solutions for $\eps=9.6\cdot 10^{-4}$
are shown in Fig.\ref{qbreatherprofile}.
The symbols correspond to normal mode energies at the moment when all
coordinates vanish.
Among them is the $q$-breather
with the strongest localization.
The precision of computation is $10^{-8}$ (see \cite{QB1} for details).
Since the normal mode energies are not conserved quantities, we show
as well the fluctuation range for each of them. These fluctuations
become stronger at particular $k$-values and are possibly
due to a closely nearby lying resonance, which nevertheless does not
destroy the localization profile.

Let us discuss the obtained results. The reason for the weaker localization of
$q$-breathers when $k_0 \gg \nu$ is the increasing distance $3k_0$
between modes excited in consecutive orders of perturbation theory.
The delocalization for $k_0 \rightarrow 0$ however is
due to an approaching of resonances $n\omega_{k_0} \rightarrow \omega_k$ for some
integer $n$ \cite{QB1}. Note that the same approaching of resonances holds at
the upper frequency cutoff where the frequency detuning is quadratic in $k$
and the relevant integer $n=1$. We computed the dependence of the slope $S$ there
and obtained a behaviour similar to the one in Fig.\ref{slopefig}.
We expect the above results to qualitatively hold independent of the dimension $D$. It remains
a challenging task to perform computations for e.g. $D=2$, since one
needs about $10^5$ lattice sites, which is presently not reachable with our
numerical tools.

We fixed the average energy density in order to ensure finite temperatures.
If the energy density $\varepsilon_{k_0}$ is fixed, then $q$-breathers
will delocalize for some nonzero value of $k_0$. For a finite lattice
$\varepsilon \approx N \varepsilon_{k_0}$ at that point.

We derived similar results for the $\alpha$-FPU model with $\phi=x+\alpha x^2$ using the estimates
from \cite{QB1}. We obtain strongest localization for $k_m\approx 2.39 \xi$
and $max(|S|) \approx 1.5/\xi$ with
the effective nonlinearity parameter $\xi=\sqrt{\alpha/\pi} \eps^{1/4}$.
For small $k_0$ the slope
$S \sim -(2k_0/\xi)^{1/3}$. 
In all cases the localization becomes meaningless when $|S|^{-1} \approx \pi$
which is the size of the first Brilloin zone. That happens at 
$\tilde{k}_0 \approx 3\beta \varepsilon/(4\pi)$ ($\beta$-FPU) and at
$\tilde{k}_0 \approx \alpha^2 \varepsilon/(2\pi^5)$ ($\alpha$-FPU). Well defined
and localized $q$-breathers exist for $k_0 \gg \tilde{k}_0$. Strong resonances
destroy them for $k_0 \ll \tilde{k}_0$ and lead to an effective redistribution of mode energy
in that hydrodynamic regime.
The same reasoning defines a critical nonlinearity value, for which $k_m$ reaches the center of the
band. It can be roughly estimated as $\nu,\xi \sim 1$.  For larger values of
$\nu,\xi$ the system will enter a regime of strong nonlinearity,
where $q$-breathers may become
meaningless.

We considered
standing waves.
We expect the results to be also of importance for
travelling waves which are reflected at boundaries or inhomogeneities.

We thank Tiziano Penati for stimulating discussions.
M.I., O.K. and K.M. appreciate the warm hospitality of the Max Planck Institute
for the Physics of Complex Systems. M.I. and O.K. also acknowledge support of
the "Dynasty" foundation.

\end{document}